\documentclass[12pt]{article}

\usepackage[totalheight = 23cm, totalwidth = 17cm]{geometry}
\usepackage{amssymb,amsmath,amsfonts,amsbsy,graphicx,bm}

\usepackage{color,cancel,ulem}

\newcommand{\dis}[1]{\begin{equation}\begin{split}#1\end{split}\end{equation}}

\begin{document}

\begin{titlepage}

\begin{center}

{\LARGE \bf 
 Bounds on tower mass scales in the presence of throats of different warping
}

\vskip 1.0cm

{\large
Min-Seok Seo$^{a}$ 
}

\vskip 0.5cm

{\it
$^{a}$Department of Physics Education, Korea National University of Education,
\\ 
Cheongju 28173, Republic of Korea
}

\vskip 1.2cm

\end{center}

\begin{abstract}

 In Type IIB flux compactification  realizing the metastable de Sitter (dS) vacuum, the uplift potential can be generated by $\overline{\rm D3}$-branes at the tip of Klebanov-Strassler throat.
 Then the uplift potential obeys the scaling law with respect to the tower mass scale  $m_{\rm sc}$, which can be  the Kaluza-Klein (KK) mass scale associated with the throat containing $\overline{\rm D3}$-branes or the bulk tower mass scales, depending on the warping of the throat.
 On the other hand, in the presence of  another throat of stronger warping, the KK mass scale associated with this throat is lower than $m_{\rm sc}$.
 Nevertheless, the Higuchi bound and the condition that the tower mass scale is higher than the gravitino mass provide the upper bound on $m_{\rm sc}$ determined by the lowest tower mass scale (or gravitino mass).
 This bound also can be interpreted as the lower bound on the lowest tower mass scale determined by $m_{\rm sc}$.
 We investigate this bound in detail when the throat containing $\overline{\rm D3}$-branes is strongly and weakly warped, respectively.

\end{abstract}

\end{titlepage}

\newpage

\section{Introduction}

The swampland program \cite{Vafa:2005ui} has provided a set of  conjectured constraints that the low energy effective field theory (EFT) must satisfy in order to have a UV completion in quantum gravity (for reviews, see, e.g., \cite{Brennan:2017rbf, Palti:2019pca, vanBeest:2021lhn, Grana:2021zvf, Agmon:2022thq, VanRiet:2023pnx}).
Among various proposals in the program, the instability of de Sitter (dS) space formulated by the dS swampland conjecture is of particular interest \cite{Dine:1985he, Danielsson:2018ztv, Obied:2018sgi} (see also \cite{Andriot:2018wzk, Garg:2018reu, Ooguri:2018wrx}) as string realization of the  metastable dS vacuum   \cite{Kachru:2003aw, Balasubramanian:2005zx} requires a tuning between several ingredients such as   flux compactification, non-perturbative effect, and uplift, which has led to the debate on the consistency of the model.
In the justification of the dS swampland conjecture  the distance conjecture plays the crucial role \cite{Ooguri:2018wrx}.
It states that the infinite distance limit of the moduli space corresponds to the corner of the landscape, at which the EFT becomes invalid as the mass scale of a tower of states decreases rapidly \cite{Ooguri:2006in}.
 Such a descent of a tower of states implies the rapid increase in the number of low energy degrees of freedom hence their production in dS space can violate the covariant entropy bound \cite{Bousso:1999xy} given by  (horizon area)$/4$  (see also \cite{Seo:2019mfk, Seo:2019wsh, Cai:2019dzj, Sun:2019obt, Seo:2023xsb}).

 The dS swampland conjecture raised the suspicion that our universe well described by  a positive   cosmological constant,  $\Lambda = 3m_{\rm Pl}^2H^2$, where $H$ is the Hubble parameter (the inverse of the horizon radius),  is close to the swampland in the moduli space.
 In this regard,  there have been attempts to formulate the closeness of our universe  to the swampland in the form of the scaling law, reflecting the distance conjecture.
  The anti-dS(AdS)/dS distance conjecture focuses on the smallness of $\Lambda\sim 10^{-123} m_{\rm Pl}^4$ and  suggests the relation $m \sim |\Lambda|^\alpha$, where $m$ is some tower mass scale  \cite{Lust:2019zwm}.
 If $m$ is the Kaluza-Klein (KK) mass scale, the lower bound on $\alpha$ is given by $1/4$ \cite{Montero:2022prj}, which can be obtained from the observational bound on the size of extra dimensions \cite{Lee:2020zjt, Hannestad:2003yd}.
 Moreover, any nonzero mass of the state  in dS space is required to be larger than the Higuchi bound 
\cite{Higuchi:1986py} (see \cite{Enayati:2022hed} for a review) given by $\sqrt{s(s-1)}H \sim \Lambda^{1/2}$, where $s$ is the spin of the state, indicating that $\alpha$ is smaller than $1/2$.
 On the other hand,  in the effective supergravity description of string theory, $|\Lambda|$ of the AdS vacuum (for the metastable dS vacuum, the size of the AdS vacuum energy density before uplift which will be denoted by $V_{\rm AdS}$) is smaller than $3m_{\rm Pl}^2 m_{3/2}^2$, where $m_{3/2}$ is the gravitino mass and the inequality is saturated in the supersymmetric case.
 The  gravitino distance conjecture claims that it may be $m_{3/2}$ rather than $|\Lambda|$ which obeys the scaling law   \cite{Palti:2020tsy, Cribiori:2021gbf, Castellano:2021yye}.

  In the string model realizing the metastable dS vacuum based on Type IIB flux compactification,   when the uplift potential $V_{\rm up}$ is generated by $\overline{\rm D3}$-branes at the tip of the Klebanov-Strassler throat \cite{Kachru:2002gs}, it turns out that the size of $V_{\rm up}$ obeys the scaling law with respect to the tower mass scale \cite{Seo:2023ssl}.
 If the throat containing $\overline{\rm D3}$-branes is strongly warped, the mass scale of the KK modes localized in this throat  region  satisfies $m_{\rm KK}^{\rm throat} \sim V_{\rm up}^{1/4}$ \cite{Blumenhagen:2022zzw}.
   We note that in the presence of a number of throats \cite{Hebecker:2006bn}, the KK mass  scale associated with the  throat of the strongest warping is the lowest tower mass scale.
   Thus, if the warping of the throat containing $\overline{\rm D3}$-branes is the strongest, the scaling law relates $V_{\rm up}$ and the lowest tower mass scale.
 Moreover, the exponent $1/4$ is nothing more than the inverse of the number of noncompact dimensions over which $\overline{\rm D3}$-branes are extended.
  In contrast, when the warping is extremely weak,  both the bulk tower mass scales and $V_{\rm up}$ scale  with respect to the size of internal volume so we can find the scaling law between them \cite{Seo:2023ssl}.
 More concretely, the string  scale $m_s$ satisfies $m_s \sim V_{\rm up}^{1/4}$ where the exponent is the inverse of the number of noncompact dimensions.
 In addition, the bulk KK mass scale, the lowest tower mass scale in the absence of the throat of   stronger warping, also obeys the scaling law but the exponent in this case is given by $1/3$ : $m_{\rm KK}^{\rm bulk}\sim V_{\rm up}^{1/3}$.

  Since the uplift potential given by $V_{\rm up}=\Lambda +|V_{\rm AdS}|$ is connected to the tower mass scale $m_{\rm sc}$ through the scaling law $m_{\rm sc} \sim V_{\rm up}^{ \alpha}$ with $\alpha=1/4$ or $1/3$, we find the inequality $m_{\rm sc}>\Lambda^\alpha$ for the positive $\Lambda$. 
 Hence, unlike the AdS/dS distance conjecture suggesting the equality, $m_{\rm sc}$ may not be as light as $\sim \Lambda^\alpha$.
 Meanwhile, if there exists a throat whose warping is stronger than that of the throat containing $\overline{\rm D3}$-branes, the KK mass scale associated with it is lower than  $m_{\rm sc}$, the tower mass scale satisfying the scaling law with respect to $V_{\rm up}$.
 Nevertheless, for the EFT based on the four-dimensional supergravity description to be valid, the lowest tower mass scale is required to be larger than $m_{3/2}$.
 Combining this with the Higuchi bound and the fact that $|V_{\rm AdS}|$ is smaller than $3m_{\rm Pl}^2 m_{3/2}^2$, we obtain the inequality obeyed by   the lowest tower mass scale as well as $m_{\rm sc}$.
 This will be explored in detail in this article.
 As we will see, in the presence of the lower tower mass scale, $m_{\rm sc}$ has the upper bound determined by the  lower tower mass scale, or equivalently, the lowest tower mass scale has the lower bound determined by $m_{\rm sc}$.
 We expect that our results may be useful in the phenomenological study on the structure of extra dimensions.
 
 This article is organized as follows.
  Section \ref{Sec:review} is devoted to the brief reviews on the Higuchi bound and the connection between tower mass scales and $V_{\rm up}$, which provide the background for our discussion.
 Based on them,  in Section \ref{Sec:bounds}, we obtain the upper bound on $m_{\rm sc}$ in terms of the  lower tower mass scale when the throat containing $\overline{\rm D3}$-branes is strongly and weakly warped, respectively.
 Then we conclude.

 \section{Reviews on Higuchi bound, tower mass scale and uplift   }
  \label{Sec:review}

 \subsection{Higuchi bound }

 We begin our review on the Higuchi bound with the discussions on the unitary irreducible representations (UIRs) of  SO(1,4) dS isometry group and their masses.
 For details, we refer the reader to \cite{Enayati:2022hed} and references therein.
  The `mass' of the state in the UIR is determined by  the quadratic Casimir,
  \dis{Q=-\frac12 L_{AB} L^{AB},}
 where $L_{AB}$ ($A, B=0, 1,\cdots, 4$) are   SO(1,4) generators, as the field equation of motion is reduced to its eigenvalue equation,
\dis{\Big(Q-\langle Q \rangle \Big) {\cal K}(x)=0.\label{eq:EoM}}
Here the field ${\cal K}(x)$   carries the tensor or spinor indices corresponding to the UIR to which the field belongs. 
  The eigenvalue of $Q$ is given by \cite{Dixmier:1961} (see also Section 12 of \cite{Enayati:2022hed})
\dis{\langle Q \rangle = -s(s+1)-(q+1)(q-2),} 
where $s$ is interpreted as a spin in the $H\to 0$ limit and  $q \in \mathbb{C}$ is determined by the type of the representation.
\footnote{A pair of numbers $(s, q)$ labelling the UIR can be obtained by observing the representation of SO(4) isometry subgroup.
Since SO(4)$\cong$ SU(2)$\times$SU(2) and   the quadratic Casimirs of two SU(2)s have eigenvalues  $j_\ell(j_\ell+1)$ and $j_r(j_r+1)$ ($j_\ell, j_r \in \mathbb{Z}/2$), respectively,    the irreducible representation of SO(4) is characterized by $(j_\ell, j_r)$.
Then $s$ is the infimum (greatest lower bound) of   $j_\ell +j_r$, which is interpreted as a spin in the $H\to 0$ limit.
On the other hand, SO(1,4)/SO(4) generators raise/lower the  $j_\ell$ and $j_r$ values, and their contributions to the SO(1,4) quadratic Casimir $Q$ determine the value of $q$.
For example,  in the discrete series of representations, we can find two dual representations in which the values of $({\rm max}(j_r-j_\ell), {\rm min}(j_r-j_\ell))$ for the states satisfying $j_r+j_\ell=s$ are given by $(q, s)$ and $(-s, -q)$,  respectively. }
The types of   SO(1,4) UIRs are classified in \cite{Dixmier:1961}, and it turns out that under the Poincar\'e contraction, i.e., in  the  $H\to 0$ limit,  a particular set of  UIRs can be reduced to the  wavefunction in   Minkowski space   in a sensible way (the positive and negative frequency modes are well separated) \cite{Garidi:2003dp} (see also Section 14 of \cite{Enayati:2022hed}) :
\begin{itemize}
\item The principal series of representations   : In this case, $q=\frac12+i\nu$, where $\nu\geq 0$ for an integer $s$ and $\nu >0$ for a half-integer $s$, such that
\dis{\langle Q\rangle= -s(s+1)+\frac94 +\nu^2.}
The  representation  of this type is also called the massive representation since in   the $H\to 0$ limit, $\nu H$ is reduced to    the mass of the field.
\item The complementary series of representations : The value of $s$ in this case is only an integer.
Moreover,  $q=\frac12+ \nu$, where $\nu \in \mathbb{R}$ and 
\dis{&0<|\nu|<\frac32\quad\quad {\rm for}\quad s=0,
\\
&0<|\nu|<\frac12\quad\quad {\rm for}\quad s=1,2,3,\cdots,}
giving
\dis{\langle Q\rangle= -s(s+1)+\frac94-\nu^2. }
The $H\to 0$ limit of the representation of this type is sensible when $s=0$ and $\nu=1/2$ (thus $q=1$), which corresponds to the conformally coupled massless spin-$0$ field.
\item The discrete series of representations : In this case, $q=0,1, \cdots, s-1, s$ for an integer $s$ and $q=\frac12, \cdots, s-1, s$ for a half-integer $s$.
The sensible representation in the $H\to 0$ limit requires that $s=q>0$, which is reduced to the massless  spin-$s$ field.
\end{itemize}
Moreover, the quadratic Casimir of dS isometry is not exactly identified with the Laplacian :  the quadratic Casimir eigenvalue equation given by \eqref{eq:EoM} is written as \cite{Garidi:2003ys}
\dis{\big(\square -H^2 s(s+2)-H^2\langle Q\rangle\big){\cal K}(x)=0.}
On the other hand, from the fact that the representation belonging to the discrete series satisfying $s=q>0$   becomes the massless spin-$s$ field in the $H\to 0$ limit, it was suggested to define the `mass' of the state in dS space by \cite{Garidi:2003ys}
\dis{m^2=H^2 \big(\langle Q\rangle-\langle Q_{s=q}\rangle\big)=H^2 \big(\langle Q\rangle+2 (s^2-1)\big)=(s-q)(s+q-1)H^2,\label{eq:dSmass}}
such that $m^2=0$ for $s=q>0$.
For the representations in the principal series, the mass defined in this way can be written as $m^2=\big(s-\frac12\big)^2H^2+ \nu^2H^2$ (note that $q=\frac12+i\nu$),  so for a finite value  of $s$, $m^2$ is reduced to $\nu^2 H^2$ in the $H \to 0$ limit.
In terms of $m$, \eqref{eq:EoM} becomes
\dis{\big(\square -(2-s(s-2))H^2 -m^2\big){\cal K}(x)=0,}
which indeed coincides with the equation of motion used in the previous literatures, e.g., \cite{Higuchi:1986py, Higuchi:1989gz, Deser:2001xr, Deser:2003gw}.

It is remarkable that apart from the issue of the sensible $H\to 0$ limit, regarding \eqref{eq:dSmass} as a mass of any state in the UIR, one finds that the nonzero value of mass has a lower bound called the Higuchi bound.
More concretely, the value of $m^2$ turns out to be either zero or larger than $s(s-1)H^2$, which is meaningful for $s>1$.
This  bound is saturated  by the representations belonging to the complementary series   with $\nu=\pm\frac12$ ($q=1,0$) and the discrete series  with $q=0$.
For instance, for $s=2$, the lower bound on the nonzero value of $m^2$ is given by $2H^2$  \cite{Higuchi:1986py} (see also \cite{Higuchi:1989gz} for $s>2$).

 \subsection{Tower mass scales in the presence of throat and uplift potential }
 
 Throughout this article, we consider  Type IIB Calabi-Yau orientifold compactifications containing a number of Klebanov-Strassler throats, in which the dilaton and complex structure moduli are stabilized by fluxes \cite{Giddings:2001yu}.
 The K\"ahler moduli are stabilized by non-perturbative effect \cite{Kachru:2003aw} and possibly  the additional $\alpha'$ corrections \cite{Balasubramanian:2005zx}, and the  potential stabilizing all the moduli is  uplifted by $\overline{\rm D3}$-branes at the tip of one of throats.
The string   scale $m_s$, the mass scale of a tower of  string excitations is given by $1/(2\pi \sqrt{\alpha'})$.
 Since the ten-dimensional gravitational coupling is given by $\kappa_{10}^2=g_s^2/(4\pi m_s^8)$, denoting the volume of the internal manifold by ${\cal V}/m_s^6$, we obtain the relation between $m_s$ and the four-dimensional Planck scale $m_{\rm Pl}$ :
  \dis{m_s=\frac{g_s}{\sqrt{4\pi{\cal V}}}m_{\rm Pl}.\label{eq:mstring}} 
 
Moreover, under the compactification, there can be various KK mass scales depending on where   the KK modes are localized.
 The mass scale of the KK modes in the bulk is given by
  \dis{m_{\rm KK}^{\rm bulk}=\frac{2\pi m_s}{{\cal V}^{1/6}}=\sqrt{\pi}\frac{g_s}{{\cal V}^{2/3}}m_{\rm Pl}.\label{eq:bulkKK}}
On the other hand, the mass scale of the KK modes localized in the throat region is determined by how strong the warping of the throat is.
 To see this, we note that the metric  near the tip of the throat is given by
  \dis{ds^2=e^{2\Omega_4(x,y)}g_{\mu\nu}dx^\mu dx^\nu + e^{2\Omega_6(x,y)}g_{mn}dx^m dx^n,}
  where 
  \dis{e^{2\Omega_4(x,y)}=e^{2A(y)}e^{2\Omega(x)},\quad\quad
  e^{2\Omega_6(x,y)}=e^{-2A(y)}\sigma(x)^{1/2}.}
   This throat geometry is supported by fluxes of $F_3$ and $H_3$, the flux quanta of which in string units will be denoted by  $M$ and $K$, respectively.
 In the metric, $\sigma(x)$ is the scalar part of the volume modulus, the vacuum expectation value of which satisfies ${\cal V}=\langle \sigma^{3/2} \rangle$ under the normalization $\int d^6 y \sqrt{g_6}e^{-4A} = m_s^{-6}$. 
 Then the Weyl factor $e^{2\Omega(x)}$ can be written as
  \dis{e^{2\Omega(x)}=\frac{{\cal V}\ell_s^6}{\sigma(x)^{3/2}\int d^6y \sqrt{g_6}e^{-4A}}=\frac{\langle \sigma^{3/2}\rangle}{\sigma(x)^{3/2}},}
  such that $\langle  e^{2\Omega(x)}\rangle=1$.
  The warping of the throat is typically parametrized by the `warp factor' defined by $e^{-4A}$, which can be written as
  \dis{e^{-4A(y)}=1+\frac{e^{-4A_0(y)}}{\sigma(x)},\label{eq:warp}}
 where
   \dis{
  &e^{-4A_0(y)}=2^{2/3}\frac{(g_sM)^2}{(2\pi)^4 |z|^{4/3}}I(\eta),\\
 & I(\eta)=\int_\eta^\infty dx \frac{x\coth x-1}{\sinh^2x}(\sinh(2x)-2x)^{1/3}.\label{eq:warp1}}
 Then the throat KK mass scale is given by $m_{\rm KK}^{\rm throat}=\langle e^{\Omega_4}\rangle /(\langle e^{\Omega_6}\rangle R)=\langle e^{2A}\rangle/ (R {\cal V}^{1/6})$, where $R$ is the typical length scale of the throat : $R\sim |z|^{1/3}/m_s$ for the A-cycle and $R\sim \eta_{\rm UV}|z|^{1/3}/m_s$ for the B-cycle, where $\eta_{\rm UV}\sim \log\big(\frac{1}{|z|}\big)=\frac{2\pi}{g_s}\frac{K}{M}$($>1$) is the length of the throat.
When the throat is strongly warped, $e^{-4A}\gg 1$ is satisfied, and the throat KK mass scale is highly redshifted by the warp factor \cite{Firouzjahi:2005qs} (see also \cite{Blumenhagen:2019qcg, ValeixoBento:2022qca} for recent discussions) :  
  \dis{m_{\rm KK}^{\rm throat}=\frac{2^{1/2}3^{1/6}\pi^{3/2}}{I(0)^{1/2}}\frac{|z|^{1/3}}{M {\cal V}^{1/3}}m_{\rm Pl}.\label{eq:KKstrong}}
  Moreover, for the KK modes localized along the B-cycle of the throat, the corresponding KK mass scale is additionally suppressed by $\eta_{\rm UV}$ \cite{Blumenhagen:2019qcg}.
  Therefore, the KK mass scale associated with the throat of the strongest warping  (that is, the smallest $|z|$ and the largest $\eta_{\rm UV}$) is typically the lowest tower mass scale.
  In contrast, when the throat is extremely weakly warped, i.e., $e^{-4A}\simeq 1$, the throat KK mass scale is given by
  \dis{m_{\rm KK}^{\rm throat}=\frac{g_s}{|z|^{1/3}{\cal V}^{2/3}}m_{\rm Pl}.}
  Comparing this with \eqref{eq:bulkKK}, one finds that for the extremely weakly warped throat, the bulk KK mass scale $m_{\rm KK}^{\rm bulk}$ is the lowest tower mass scale.

 Now we move onto the uplift potential $V_{\rm up}$.
 When ${\overline {\rm D3}}$-branes at the tip of the throat are extended over the noncompact four-dimensional spacetime,   the induced metric is given by
 \dis{ds_{{\overline {\rm D3}}}^2=e^{2\Omega_4(x,y)}g_{\mu\nu}dx^\mu dx^\nu = e^{2A(y)}e^{2\Omega(x)} g_{\mu\nu}dx^\mu dx^\nu,}
 from which $V_{\rm up}$ is written as
 \dis{V_{\rm up}=2p \frac{T_3}{g_s}e^{4\Omega_4(x,y)} = 4\pi p \frac{m_s^4}{g_s}e^{4A(y)}e^{4\Omega(x)},}
 where $T_3=2\pi m_s^{4}$ is the brane tension and $p$ is the number of $\overline{\rm D3}$-branes.
 For the strongly warped throat ($e^{-4A}\gg 1$),  we obtain
  \dis{V_{\rm up} =\frac{2^{4/3}\pi^3}{I(0)}\frac{g_sp}{M^2}\frac{|z|^{4/3}}{{\cal V}^{4/3}}m_{\rm Pl}^4.}
  Comparing this with \eqref{eq:KKstrong}, one finds the scaling law,
  \dis{m_{\rm KK}^{\rm throat}\sim \frac{1}{g_s^{1/4}M^{1/2} p^{1/4}}\langle V_{\rm up}\rangle^{1/4}, \label{eq:scLstrong}}
  where  the exponent $1/4$ is the inverse of the number of noncompact dimensions over which $\overline{\rm D3}$-branes are extended.
  Meanwhile, when the throat is weakly warped ($e^{-4A}\simeq 1$),  the uplift potential is given by
   \dis{V_{\rm up} =\frac{g_s^3}{4\pi}\frac{p}{{\cal V}^2}m_{\rm Pl}^4,}
   from which one finds two scaling laws with respect to $m_s$ and $m_{\rm KK}^{\rm bulk}$ given by \eqref{eq:mstring} and \eqref{eq:bulkKK}, respectively : 
 \dis{& m_s \sim \Big(\frac{g_s}{4\pi p}\Big)^{1/4}\langle V_{\rm up}\rangle^{1/4},
 \\
 &m_{\rm KK}^{\rm bulk} \sim \frac{1}{p^{1/3}}\Big\langle\frac{V_{\rm up}}{m_{\rm Pl}^4}\Big\rangle^{1/3} m_{\rm Pl}.\label{eq:scLweak}}
 We note that the exponent in the scaling law with respect to $m_s$ is $1/4$, the inverse of the number of noncompact dimensions.
 Moreover,  $m_{\rm KK}^{\rm bulk}$ is the lowest tower mass scale.

\section{Relation between two tower mass scales}
\label{Sec:bounds}

 In string model, the metastable dS vacuum   is realized by uplift of the AdS vacuum, indicating that  the positive cosmological constant $\Lambda$ can be written as
 \dis{\Lambda= 3m_{\rm Pl}^2H^2=-|V_{\rm AdS}|+V_{\rm up}.\label{eq:master}}
 As reviewed in the previous section, $V_{\rm up}$ obeys the scaling law with respect to some tower mass scale which will be denoted by $m_{\rm sc}$,
  \dis{V_{\rm up}= v_0 m_{\rm Pl}^4\Big(\frac{m_{\rm sc}}{m_{\rm Pl}}\Big)^{1/\alpha}.} 
 For the strongly warped throat, $m_{\rm sc}$ is identified with $m_{\rm KK}^{\rm throat}$ given by \eqref{eq:KKstrong} and $\alpha=1/4$.
 For the extremely weakly warped  throat, $m_{\rm sc}$ corresponds to $m_s$ given by \eqref{eq:mstring} ($\alpha=1/4$) or $m_{\rm KK}^{\rm bulk}$ given by \eqref{eq:bulkKK} ($\alpha=1/3$).

 We now consider   another throat of  stronger warping such that the mass scale $m_0$ of KK modes localized in this throat is lower than $m_{\rm sc}$. 
 Denoting the conifold modulus and the flux quanta of $F_3$ associated with the throat of the stronger warping by $z_\ell$ and $M_\ell$, respectively, we obtain
 \dis{m_0 \simeq \frac{|z_\ell |^{1/3}}{M_\ell {\cal V}^{1/3}}m_{\rm Pl},\label{eq:longerKK}}
 just like \eqref{eq:KKstrong}.
  The Higuchi bound imposes that the nonzero $m_0$ is larger than $\sqrt{s(s-1)}H$, or equivalently, $H<m_0/\sqrt{s(s-1)}$.
  Combining this with the fact that $|V_{\rm Ads}|$ is  smaller than $3m_{\rm Pl}^2 m_{3/2}^2$, \eqref{eq:master} provides the inequality
\dis{3m_{\rm Pl}^2 m_{3/2}^2 >|V_{\rm AdS}| > v_0 m_{\rm Pl}^4\Big(\frac{m_{\rm sc}}{m_{\rm Pl}}\Big)^{1/\alpha} - \frac{3}{s(s-1)} m_{\rm Pl}^2 m_0^2,\label{eq:ineqA}}
which relates three   scales, $m_{\rm sc}$, $m_0$, and $m_{3/2}$.
 Noting that   $m_{3/2}$ is the characteristic mass scale of the four dimensional supergravity formalism,   we expect that this is lower than $m_0$, the KK mass scale implying the existence of extra dimensions, hence $3m_{\rm Pl}^2 m_0^2>3 m_{\rm Pl}^2 m_{3/2}^2$.
 Then we obtain the inequality
 \dis{3\Big(1+\frac{1}{ s(s-1)}\Big)m_{\rm Pl}^2 m_0^2 > v_0 m_{\rm Pl}^4\Big(\frac{m_{\rm sc}}{m_{\rm Pl}}\Big)^{1/\alpha}, \label{eq:ineqB}}
 which indicates that $m_{\rm sc}$ ($m_0$) has the upper (lower) bound determined by $m_0$ ($m_{\rm sc}$).
   We investigate this inequality in detail when the throat containing $\overline{\rm D3}$-branes is strongly and weakly warped, respectively.

\subsection{Strong warping case ($e^{-4A} \gg 1$)}

 When the throat containing $\overline{\rm D3}$-branes is strongly warped, the scaling law  \eqref{eq:scLstrong} is satisfied,   indicating   $m_{\rm sc}=m_{\rm KK}^{\rm throat}$ (given by \eqref{eq:KKstrong}), $\alpha=1/4$, and $v_0\simeq g_s p M^2$.
  Then  \eqref{eq:ineqA} is written as
\dis{3m_{\rm Pl}^2 m_{3/2}^2 >|V_{\rm AdS}| > (g_s p M^2) (m_{\rm KK}^{\rm throat})^4 - \frac{3}{s(s-1)} m_{\rm Pl}^2 m_0^2,\label{eq:Ineq0}}
and  \eqref{eq:ineqB} becomes
 \dis{m_{\rm KK}^{\rm throat}<\Big(\frac{3(s^2-s+1)}{(g_s p M^2) s(s-1)}\Big)^{1/4} m_{\rm Pl}^{1/2} m_0^{1/2},\label{eq:Ineq1}}
  or equivalently,
 \dis{\Big(\frac{m_0}{m_{\rm KK}^{\rm throat}}\Big)^2>(g_s p M^2)\frac{s(s-1)}{3(s^2-s+1)}\Big(\frac{m_{\rm KK}^{\rm throat}}{m_{\rm Pl}}\Big)^2.}
  We may rewrite this bound in terms of the explicit expressions \eqref{eq:KKstrong} for $m_{\rm KK}^{\rm throat}$ and \eqref{eq:longerKK} for $m_0$ to obtain the   constraint  on the warping of throats :
  \dis{\Big(\frac{|z_\ell|}{|z|}\Big)^{1/3}\frac{M}{M_\ell} > \Big(\frac{(g_s p M^2) s(s-1)}{3(s^2-s+1)}\Big)^{1/2} \frac{|z|^{1/3}}{M {\cal V}^{1/3}}.} 
The bound \eqref{eq:Ineq1} shows that even if the throat containing $\overline{\rm D3}$-branes is not of the strongest warping so the associated KK mass scale is not the lowest tower mass scale, it cannot be arbitrarily higher than the lowest KK mass scale $m_0$, but bounded by $\lesssim m_{\rm Pl}^{1/2} m_0^{1/2}$.
This in turn means that $m_0$ cannot be arbitrarily small, but has the lower bound determined by $m_{\rm KK}^{\rm throat}$.
 Indeed, the size of deformation at the tip of the throat is given by $(g_sM)^{1/2}/(2\pi m_s)$, 
 \footnote{This comes from the fact that for the strongly warped throat, $e^{-2A}$ is approximated by $e^{-2A_0}/\sigma^{1/2}$ (see \eqref{eq:warp}).
 As can be inferred from \eqref{eq:warp1}, it  depends on $|z|$ and $\sigma$ through the combination $1/(|z|^{2/3}\sigma^{1/2})$, which is cancelled by the prefactor $\sigma^{1/2}$ in $G_{mn}=e^{-2A}\sigma^{1/2}g_{mn}$ and the overall factor $|z|^{2/3}$ in $g_{mn}$ \cite{Klebanov:2000hb}.
 As a result, the overall factor in $G_{mn}$ is given by $(g_sM)/(2\pi m_s)^2$.
  }
which is required to be much larger than the string length  $m_s^{-1}$ for the effective supergravity description to be valid. 
This indicates that $g_s M \gg 1$.
Moreover, we can further impose $g_sM^2 \gg p$ because otherwise the conifold modulus $z$ is stabilized at $0$  \cite{Bena:2018fqc} (see, however,   \cite{Lust:2022xoq} for a recent counterargument).
Both of these constraints impose that $g_s p M^2 \gg 1$, so our bound is more or less stronger than the simple inequality $m_{\rm KK}^{\rm throat}< {\cal O}(1) m_{\rm Pl}^{1/2} m_0^{1/2}$.

 In fact,  \eqref{eq:Ineq1} as the upper bound on $m_{\rm KK}^{\rm throat}$ is useful when the size of $\Lambda$ is not negligibly small, which may be realized in the inflationary cosmology.
 In contrast, $\Lambda$ in our universe is as small as $10^{-123}m_{\rm Pl}^4$ so it is reasonable to take $|V_{\rm up}|$ to be much larger than $\Lambda$.
 Since  $V_{\rm up}\simeq |V_{\rm AdS}|$ in this case, the condition $|V_{\rm AdS}|<3m_{\rm Pl}^2 m_{3/2}^2$ gives a more stringent bound,
\dis{m_{\rm KK}^{\rm throat} \lesssim \Big(\frac{3}{g_s pM^2}\Big)^{1/4} m_{\rm Pl}^{1/2} m_{3/2}^{1/2}.}
But still, the bound \eqref{eq:Ineq1} is valid as the lower bound on $m_0$ determined by $m_{\rm KK}^{\rm throat}$.
In any case, it is remarkable that some intermediate new physics scale has the upper bound   determined by another much lower   scale.

 We can also compare the bound \eqref{eq:Ineq1} with the bound obtained from the species scale $\Lambda_{\rm sp}$ \cite{Dvali:2007hz, Dvali:2007wp},
 \dis{\Lambda_{\rm sp}=\frac{m_{\rm Pl}}{\sqrt{N_{\rm tot}}},}
 above which gravity is no longer weakly coupled to matter.
 Here $N_{\rm tot}$ is the number of low energy degrees of freedom below $\Lambda_{\rm sp}$, hence given by 
 \dis{&N_{\rm tot}=N_0+N_{\rm KK},
 \\
 &N_0=\frac{\Lambda_{\rm sp}}{m_0},\quad\quad N_{\rm KK}=\frac{\Lambda_{\rm sp}}{m_{\rm KK}}.}
For $m_0 \ll m_{\rm KK}$, we have $N_0 \gg N_{\rm KK}$ then $\Lambda_{\rm sp}\simeq m_{\rm Pl}/\sqrt{N_0}$, from which we obtain $\Lambda_{\rm sp} \simeq m_{\rm Pl}^{2/3}m_0^{1/3}$.
Since  $m_{\rm KK}<\Lambda_{\rm sp}$, a condition $m_{\rm KK}<m_{\rm Pl}^{2/3}m_0^{1/3}$ is satisfied, but this is less stringent than  \eqref{eq:Ineq1}.

\subsection{Weak warping case ($e^{-4A} \simeq 1$)}

 When the throat containing $\overline{\rm D3}$-branes is extremely weakly  warped, two   scaling laws given by \eqref{eq:scLweak} are satisfied :   for $m_{\rm sc}=m_{\rm KK}^{\rm bulk}$ (given by \eqref{eq:bulkKK})  $v_0\simeq p$ and $\alpha=1/3$ while  for $m_{\rm sc}=m_s$ (given by \eqref{eq:mstring}) $v_0\simeq (4\pi p) /g_s$ and $\alpha=1/4$.

  We first consider the case $m_{\rm sc}=m_{\rm KK}^{\rm bulk}$, in which the inequality \eqref{eq:ineqA}  is written as 
\dis{3m_{\rm Pl}^2 m_{3/2}^2 >|V_{\rm AdS}| > p  m_{\rm Pl} (m_{\rm KK}^{\rm bulk})^3 - \frac{3}{s(s-1)} m_{\rm Pl}^2 m_0^2.}
In the presence of an additional throat which is strongly warped,   the KK modes localized in this throat provide the lowest tower mass scale given by \eqref{eq:longerKK}.
Requiring $m_0>m_{3/2}$, we obtain
 \dis{3\Big(1+\frac{1}{ s(s-1)}\Big)m_{\rm Pl}^2 m_0^2 >  p  m_{\rm Pl}(m_{\rm KK}^{\rm bulk})^3,}
 or equivalently,
 \dis{m_{\rm KK}^{\rm bulk}<\Big(\frac{3(s^2-s+1)}{p s(s-1)}\Big)^{1/3} m_{\rm Pl}^{1/3} m_0^{2/3}.}
 Thus, $m_{\rm KK}^{\rm bulk}$ has the upper bound depending on $m_0$, with the exponent given by $2/3$.
At the same time, this inequality may be interpreted as the lower bound on $m_0$ as well.
 Putting the explicit expressions for $m_0$ and $m_{\rm KK}^{\rm bulk}$ into the inequality gives the constraint on the strongest warp factor   : 
 \dis{\frac{3(s^2-s+1)}{p s(s-1)} \Big(\frac{|z_\ell|^{1/3} }{  M_\ell}\Big)^2> \frac{g_s^3}{{\cal V}^{4/3}}.}
 Just like the previous case in which the throat containing $\overline{\rm D3}$-branes is strongly warped, we have a more stringent upper bound on $m_{\rm KK}^{\rm bulk}$ determined by $m_{3/2}$ if $|V_{\rm AdS}|\simeq V_{\rm up} \gg \Lambda$,
 \dis{m_{\rm KK}^{\rm bulk} <\frac{3}{p} m_{\rm Pl}^{1/3} m_{3/2}^{2/3}.}

 On the other hand, when $m_{\rm sc}=m_s$, the  inequality \eqref{eq:ineqA}  is written as 
\dis{3m_{\rm Pl}^2 m_{3/2}^2 >|V_{\rm AdS}| > \frac{4\pi p}{g_s}m_s^4   - \frac{3}{s(s-1)} m_{\rm Pl}^2 m_0^2,\label{eq:msraw}}
and   \eqref{eq:ineqB} reads
  \dis{m_s<\Big(\frac{3 g_s(s^2-s+1)}{(4\pi p) s(s-1)}\Big)^{1/4} m_{\rm Pl}^{1/2} m_0^{1/2}.\label{eq:msbound}}
Putting an explicit expression for $m_0$ given by \eqref{eq:longerKK} into the inequality, we obtain the   bound on the strongest warp factor : 
 \dis{\frac{m_s}{m_{\rm Pl}}<\Big(\frac{3 g_s(s^2-s+1)}{(4\pi  p  ) s(s-1)}\Big)^{1/4}  \Big(\frac{m_0}{m_{\rm pl}}\Big)^{1/2} = \Big(\frac{3 g_s(s^2-s+1)}{(4\pi  p  ) s(s-1)}\Big)^{1/4} \Big(\frac{|z_\ell |^{1/6}}{M_\ell^{1/2} {\cal V}^{1/6}}\Big).  \label{eq:msmpl} } 
For $|V_{\rm AdS}|\simeq V_{\rm up} \gg \Lambda$, we have a more stringent bound on  $m_s$,
\dis{m_s < \Big(\frac{3g_s}{4\pi p}\Big)^{1/4} m_{\rm Pl}^{1/2} m_{3/2}^{1/2}.}

 We can also compare our lower bound on $m_s$ given by \eqref{eq:msbound} with the lower bound on $m_s$ considered in \cite{Noumi:2019ohm, Lust:2019lmq}.
 This comes from the observation that the mass and spin of string excitations satisfy the Regge trajectory relation,
 \dis{m^2 = (s-1) m_s^2,\label{eq:Regge}}
or $m^2\simeq s m_s^2$ for large $s$. 
 This relation, however, violates the Higuchi bound $m^2 > s(s-1)H^2 \simeq s^2H^2$ when the spin is larger than $s_{\rm max}= (m_s/H)^2$, which implies that  the cutoff scale is lower than   $\sqrt{s_{\rm max}}m_s = m_s^2/H$.  
 If we identify the cutoff scale  with $m_{\rm Pl}$, we obtain the inequality $m_s > H^{1/2}m_{\rm Pl}^{1/2}$.
  Combining this with \eqref{eq:msmpl}, we obtain the bound 
 \dis{\frac{H}{m_{\rm Pl}} <\Big(\frac{3 g_s(s^2-s+1)}{(4\pi  p  ) s(s-1)}\Big)^{1/2}  \frac{m_0}{ m_{\rm Pl}}  = \Big(\frac{3 g_s(s^2-s+1)}{(4\pi  p  ) s(s-1)}\Big)^{1/2} \Big(\frac{|z_\ell |^{1/3}}{M_\ell {\cal V}^{1/3}}\Big),} 
or roughly, $H<m_0$, which is more or less equivalent to the Higuchi bound.
 The similar conclusion can be drawn by combining 
 \dis{3 m_{\rm Pl}^2 m_0^2> 3m_{\rm Pl}^2 m_{3/2}^2 >|V_{\rm AdS}|=\frac{4\pi p}{g_s}m_s^4-3 m_{\rm Pl}^2 H^2\label{eq:msraw2}}
  with  $m_s>H^{1/2}m_{\rm Pl}^{1/2}$ :
 \dis{3m_{\rm Pl}^2m_0^2 > \Big(\frac{4\pi p}{g_s}-3\Big)m_{\rm Pl}^2H^2.}
Meanwhile, if the cutoff scale is given by the species scale $\Lambda_{\rm sp}\simeq m_{\rm Pl}^{2/3}m_0^{1/3}$, we obtain  $m_s > H^{1/2}\Lambda_{\rm sp}^{1/2} > H^{1/2} m_{\rm Pl}^{1/3}m_0^{1/6} $.
Combining this with \eqref{eq:msmpl} gives 
 \dis{\frac{H m_0^{1/3}}{m_{\rm Pl}^{4/3}}< \Big(\frac{3 g_s(s^2-s+1)}{(4\pi  p  ) s(s-1)}\Big)^{1/2}  \Big(\frac{m_0}{m_{\rm pl}}\Big),}
 or roughly, $m_0>(H/m_{\rm Pl})^{3/2}m_{\rm Pl}$, while with \eqref{eq:msraw2} gives the trivial bound 
\dis{3m_{\rm Pl}^2m_0^2 >3 m_{\rm Pl}^2m_{3/2}^2 > \frac{4\pi p}{g_s}m_s^4-3m_{\rm Pl}^2H^2 > \frac{4\pi p}{g_s}H^2 m_{\rm Pl}^{4/3}m_0^{2/3}-3m_{\rm Pl}^2H^2,}
 since the rightmost term is negative.

   We close this section by pointing out that   the spin $s$ in the Regge trajectory relation \eqref{eq:Regge} is  identified with the level of string excitations in the Minkowski background.
   Moreover, the massive field in Minkowski space is obtained by the Poincar\'e contraction (taking $H\to 0$ limit) of the representation in the principal series in which the squared dS mass given by $\big(s-\frac12\big)^2H^2+ \nu^2H^2$.
   Since   $\nu/H$ is the mass of the field in Minkowski space, one may be tempted to identify $\nu/H$, rather than the dS mass,  with the mass in \eqref{eq:Regge},  $\sqrt{s-1} m_s$.
   In this case, the additional term $(s-\frac12)^2 H^2$ in the squared dS mass is regarded as the effect of interaction with the background geometry.
   Then  the condition $s m_s^2 > s^2 H^2$ can be interpreted as follows.
  So far as the model for dS space based on the four-dimensional particle description is concerned, $m_s$ as well as $m_{\rm KK}$ is  larger than $H$.
  That is, for the string excitations, $\nu^2 H^2 \simeq s m_s^2$ in the squared dS mass is typically   dominant over $(s-\frac12)^2H^2$ such that if $H \ll m_s$, the dS mass is approximated by the mass in the Minkowski background.
  This approximation breaks down when $s >s_{\rm max}\simeq (m_s/H)^2$, i.e., the condition $s m_s^2 > s^2 H^2$ is violated.
  In this case, the dS mass can be approximated by $(s-\frac12)H$ implying that  neglecting $H$ is no longer a good approximation even if $H \ll m_s$.

 \section{Conclusions}
\label{sec:conclusion}

 In this article,  we investigate the particular case of Type IIB orientifold compactification with fluxes in which the internal manifold contains a number of throats and the warping of the throat containing $\overline{\rm D3}$-branes is not the strongest.
 Then the tower mass scale $m_{\rm sc}$ satisfying the scaling law with respect to the uplift potential generated by $\overline{\rm D3}$-branes is not the lowest, but has the upper bound determined by the lowest tower mass scale $m_0$, typically given by the KK mass scale associated with the throat of the strongest warping.
 This also may be interpreted as the lower bound on $m_0$ determined by $m_{\rm sc}$.
 When the exponent $\alpha$ in the scaling law $m_{\rm sc}\sim V_{\rm up}^\alpha$ is $1/4$, the inverse of the number of noncompact spacetime dimensions over which $\overline{\rm D3}$-branes are extended, the upper bound on $m_{\rm sc}$ is given by $\sim m_{\rm Pl}^{1/2}  m_0^{1/2}$.
 This shows that if $m_0$ is about $10$ TeV, just above the scale accessible at the LHC search, $m_{\rm sc}$ cannot be higher than the intermediated scale, $\sim 10^{11}$ GeV.
 This bound is applied to $m_{\rm sc}=m_{\rm KK}^{\rm throat}$ when the throat containing $\overline{\rm D3}$-branes is strongly warped and $m_{\rm sc}=m_s$ when the throat containing $\overline{\rm D3}$-branes is extremely weakly warped.
 On the other hand, when the throat containing $\overline{\rm D3}$-branes is extremely weakly warped, $\alpha=1/3$ is allowed for $m_{\rm sc}=m_{\rm KK}^{\rm bulk}$.
 In this case, the upper bound on $m_{\rm sc}$ is given by $m_{\rm Pl}^{1/3}m_0^{2/3}$ which is about  $5\times 10^8$ GeV when $m_0\simeq 10$ TeV.
 We also point out that  the cosmological constant in our universe can be much smaller than the uplift potential, which allows the stronger upper  bound on $m_{\rm sc}$ in which $m_0$ is replaced by the lower scale,  $m_{3/2}$.
These bounds tell us how the structure of the internal manifold is reflected in the relations between different tower mass scales.
In particular, our setup in which   the internal manifold contains a number of throats and the warping of the throat associated with uplift is not the strongest   predicts that the evidences of the extra dimensions as well as the string may be found under the intermediate scale depending on the value of $m_0$.

%

%


\appendix



\renewcommand{\theequation}{\Alph{section}.\arabic{equation}}


\subsection*{Acknowledgements} This work is motivated by Eran Palti's question about the author's parallel session talk at SUSY 2023 conference.
This work was supported by the National Research Foundation of Korea (NRF) grant funded by the Korea government (MSIT) (2021R1A4A5031460).

\end{document}